\documentclass[twocolumn]{aastex6}		%% which will produce a one-column, single-spaced document
\pdfoutput=1
\usepackage{natbib}

\shorttitle{H$\alpha$ emission and photometric variability are not correlated in L0--T8 dwarfs}
\shortauthors{Miles-P\'aez et al.}

\begin{document}

\title{Weather on Other Worlds. IV.   H$\alpha$ emission and photometric variability are not correlated in L0--T8 dwarfs}

\author{Paulo~A.~Miles-P\'aez\altaffilmark{1}, Stanimir~A.~Metchev\altaffilmark{1,2}, Aren~Heinze\altaffilmark{3}} \and \author{D\'aniel Apai\altaffilmark{4,5}}

\altaffiltext{1}{Department of Physics \& Astronomy and Centre for Planetary Science and Exploration, The University of Western Ontario, London, Ontario N6A 3K7, Canada, (\texttt{ppaez@uwo.ca})}
\altaffiltext{2}{Department of Physics \& Astronomy, Stony Brook University, Stony Brook, New York 11743, USA}
\altaffiltext{3}{Institute for Astronomy, University of Hawaii, 2680 Woodlawn Drive, Honolulu, HI 96822, USA}
\altaffiltext{4}{The University of Arizona, Department of Astronomy, 933 North Cherry Avenue, Tucson, AZ 85721, USA}
\altaffiltext{5}{The University of Arizona, Department of Planetary Sciences and Lunar and Planetary Laboratory, 1629 East University Boulevard, Tucson, AZ 85721, USA}

\begin{abstract}
Recent photometric studies have revealed that surface spots that produce flux variations are present on virtually all L and T dwarfs. Their likely magnetic or dusty nature has been a much-debated problem, the resolution to which has been hindered by paucity of diagnostic multi-wavelength observations.  To test for a correlation between magnetic activity and photometric variability, we searched for H$\alpha$ emission among eight L3--T2 ultra-cool dwarfs with extensive previous photometric monitoring, some of which are known to be variable at 3.6 $\mu$m or 4.5 $\mu$m. We detected H$\alpha$ only in the non-variable T2 dwarf 2MASS J12545393$-$0122474. The remaining seven objects do not show H$\alpha$ emission, even though six of them are known to vary photometrically. Combining our results with those for 86 other L and T dwarfs from the literature show that the detection rate of H$\alpha$ emission is very high (94\%) for spectral types between L0 and L3.5 and much smaller (20\%) for spectral types $\ge$L4, while the detection rate of photometric variability is approximately constant (30\%--55\%) from L0 to T8 dwarfs. We conclude that chromospheric activity, as evidenced by H$\alpha$ emission, and large-amplitude photometric variability are not correlated.  Consequently, dust clouds are the dominant driver of the observed variability of ultra-cool dwarfs at spectral types at least as early as L0.

\end{abstract}

\keywords{brown dwarfs --- stars: activity --- stars: low-mass ---  stars: rotation --- stars: variables }

\section{Introduction}
\label{sec:intro}

Observational studies have shown that ultra-cool dwarfs (spectral types $\ge$M7) can display spectro-photometric variability with typical timescales of a few hours  \citep{2001AA...367..218B,2002ApJ...577..433G,2008MNRAS.391L..88L,2009ApJ...701.1534A,2012ApJ...750..105R,2013ApJ...768..121A,2013MNRAS.428.2824K,2014ApJ...793...75R,2014Natur.505..654C,2016ApJ...826....8Y}.  Accurate spectroscopic time series using the \textit{Hubble Space Telescope} and precise 3--5 $\mu$m monitoring with the \textit{Spitzer Space Telescope}, as part of the \textit{Storms} and \textit{Weather on Other Worlds} campaigns, indicate  that atmospheric spots responsible for this variability are ubiquitous on L3--T8 dwarfs \citep{2014ApJ...782...77B,2015ApJ...799..154M}. Some of these observations, combined with state-of-the-art radiative transfer models, have shown that the detected spots can generally be attributed to atmospheric dust clouds that modulate the object's brightness as it rotates   \citep{2009ApJ...701.1534A,2010ApJ...723L.117M,2012ApJ...750..105R,2012ApJ...756..172M,2013ApJ...768..121A,2015ApJ...798L..13Y}. However, the consistent detection of variability in the warmest spectral types of ultra-cool dwarfs (late-M and early L) also points to another possibility: magnetically induced chromospheric activity and hot or cold star spots, similarly to F-M stars, for which magnetic activity can produce spots \citep{1994A&A...281..395S} that lead to photometric variability (e.g., \citealt{1990ApJS...74..225H,1995AJ....110.2926H}, or \citealt{1997A&AS..125...11S}).

Magnetic activity can be revealed in different ways, for example, as emission lines from the chromosphere (e.g., \ion{Ca}{2} H and K lines) or from the transition region  (e.g., \ion{C}{4}), coronal X-rays, radio emission, and/or spots and flares \citep[][and references therein]{2013pss4.book..337R}. In most cases the observation of these activity indicators is extremely challenging for ultra-cool dwarfs: observations at X-rays and radio wavelengths are limited to the closest ultra-cool dwarfs; while \ion{Ca}{2} H and K emissions ($\lambda$3968, $\lambda$3933 \AA) are almost undetectable given the low fluxes of ultra-cool dwarfs at $\le$6000\,\AA. Several works on F-M stars have shown that H$\alpha$ emission is also an appropriate tracer of magnetic activity as it correlates with the \ion{Ca}{2} H and K lines \citep{1983ApJS...53..815Z,1991A&A...251..199P,1995A&A...294..165M}, the \ion{C}{4} line \citep{1991A&A...252..203R}, and/or X-rays emission \citep{1989A&A...218..195D,1989ApJ...344..427Y}. H$\alpha$ emission is one of the most suitable activity indicators in ultra-cool dwarfs, which emit most of their flux at red optical and near-infrared wavelengths. Observations of H$\alpha$ emission have revealed that virtually all SDSS late-M dwarfs are chromospherically active \citep{2007AJ....133.2258S,2011AJ....141...97W}. While for L dwarfs, \citet{2007AJ....133.2258S,2015AJ....149..158S} observed a decreasing fraction of H$\alpha$ emitters into the L€™s; and \citet{2016ApJ...826...73P} found that  9.2$\pm^{3.5}_{2.1}\,\%$ of L4--T8 dwarfs show H$\alpha$ emission. Star spot-like cool regions with frozen-in magnetic field lines are not expected in the highly neutral atmospheres of ultra-cool dwarfs \citep{2003ApJ...583..451M}. However, energetic magnetic field discharges ($\sim$10 kG) are a potential source of chromospheric heating (``hot spots'') as seen in some late-M and early-L dwarfs \citep[e.g.,][]{2002ApJ...572..503B,2006ApJ...648..629B,2008ApJ...673.1080B,2011A&A...525A..39Y}. More recently, optical and radio aurorae have also been proposed as likely drivers of some of the observed photometric variability at optical and near-infrared wavelengths \citep{2015Natur.523..568H,2016ApJ...818...24K}.

The likely magnetic or dusty nature of the atmospheric inhomogeneities of ultra-cool dwarfs has been a much-debated problem almost since the discovery of the first brown dwarfs \citep{1999MNRAS.304..119T,2001ApJ...557..822M,2002A&A...389..963B,2007ApJ...668L.163L}. Recent multi-wavelength photometric studies in early L dwarfs have shown sinusoidal periodicities that are well-explained by a thick long-lived cloud. \citet{2013ApJ...767..173H} presented data for the L3 dwarf DENIS-P J1058.7$-$1548 in the $J$, 3.6 $\mu$m, and  4.5 $\mu$m bands, and showed that it is difficult to account for the observed photometric variability by magnetic phenomena unless they are combined with cloud inhomogeneities. Also, \citet{2015ApJ...813..104G} showed that the optical light curve of the L1 dwarf WISEP J190648.47$+$401106.8 has remained stable for $\sim$2 years and in phase with other light curves at 3.6 $\mu$m and 4.5 $\mu$m, while its variable H$\alpha$ emission is not synchronized with the light curves. Such observational evidence points to the existence of a long-lived cloud.

Nonetheless, with chromospheric activity expected to produce star spot-like inhomogeneities, it is important to address the question whether star spots, rather than dust clouds, may dominate the photometric variability of active ultra-cool dwarfs.  We address this problem through a combination of deep optical spectroscopy of a sample of eight ultra-cool dwarfs, whose photometric light curves have been studied to a high precision with \textit{Spitzer}, and a literature sample of 86 photometrically and spectroscopically observed L0-T8 dwarfs. We seek to answer whether the combined sample shows a correlation between magnetic activity---probed by the detection of H$\alpha$ emission---and photometric variability. We briefly describe our sample in Section \ref{sec:sample} and the observations and data reduction in Section \ref{sec:obs}. The main results of our survey and its combination with literature data are presented in Sections \ref{sec:observed_sample}--\ref{sec:literature_sample}.  We discuss their implications for the correlation between observed chromospheric activity and photometric variability in Section~\ref{sec:discussion}, and summarize our findings in Section \ref{sec:conclusion}.

\section{Sample selection}
\label{sec:sample}

We selected seven L3--L6 dwarfs and one T2 brown dwarf from the sample of \citet{2015ApJ...799..154M}. Our targets are among the brightest ($J$$\sim$13.2--16.3 mag) dwarfs studied in that work, in which they were continuously monitored in the 3.6 $\mu$m and 4.5 $\mu$m bands for a total of 16--21 h by using the IRAC instrument on {\sl Spitzer}. 

Five objects exhibited photometric periodicities in the  2.7--19 h range, attributed to rotation. Another of our targets seems to be a long periodicity variable with a time scale greater than 50 h, and the remaining two did not show any photometric variability within $\pm$0.91\% in either of the 3.6 $\mu$m and 4.5 $\mu$m bands. 

The full names of our targets, their spectral types, variability period (if measured), and $J$-band magnitudes are listed in columns 1--4 of Table \ref{tab1}. Henceforth we will use abridged names for the targets. For more details about the objects we refer to \citet{2015ApJ...799..154M}.

%\floattable
\begin{deluxetable*}{lcccccccc}
\tablecaption{Targets information and observing log \label{tab1}}
\tablecolumns{10}
\tablenum{1}
\tablewidth{0pt}
\tabletypesize{\scriptsize}
\colnumbers
\tablehead{
\colhead{Object} &
\colhead{SpT\tablenotemark{a}} &
\colhead{Var. period\tablenotemark{a} (h)} &
\colhead{$J$ (mag)} &
\colhead{UT Date} &
\colhead{Inst.} &
\colhead{Config.} &
\colhead{t$_{\rm exp}$\tablenotemark{b} (s)} & 
\colhead{Air Mass} 
}
\startdata
2MASS J01033203$+$1935361	& L6	& $2.7\pm0.1$	&$16.29\pm0.08$	& 2013 Sep 02 	& GMOS-N		& Blue	& 6$\times$900 & 1.06$-$1.00  \\
							& 	& 				&	& 2013 Sep 29 	& GMOS-N		& Blue	& 10$\times$900 & 1.00$-$1.11  \\
2MASS J11263991$-$5003550	& L4.5& $3.2\pm0.3$	&$14.00\pm0.03$	& 2014 Mar 05 	& GMOS-S		& Red	& 8$\times$900  & 1.12$-$1.08 \\ 
							& 	&				&	& 2014 Mar 07 	& GMOS-S		& Red	& 8$\times$900  & 1.07$-$1.25 \\ 
2MASS J12545393$-$0122474	& T2	& \nodata	 		&$14.89\pm0.04$	& 2014 Mar 09 	& GMOS-S		& Red	& 6$\times$925 & 1.14$-$1.19  \\
							& 	&				&	& 2014 May 04 	& GMOS-S		& Red	& 8$\times$925 & 1.30$-$1.14  \\
							&  	& 		 		&	& 2014 May 05 	& GMOS-S		& Red	& 8$\times$925 & 1.24$-$1.14  \\
2MASS J14162408$+$1348263	& L6	& \nodata			&$13.15\pm0.03$	& 2014 Apr 23		& GMOS-N		& Blue	& 8$\times$780 & 1.07$-$1.40 \\
2MASS J17210390$+$3344160	& L3	& $2.6\pm0.1$	&$13.63\pm0.02$	& 2014 May 01 	& GMOS-N		& Blue	& 6$\times$900 & 1.06$-$1.03 \\
							& 	& 				&	& 2014 May 02 	& GMOS-N		& Blue	& 1$\times$900 & 1.04$-$1.05 \\
							& 	& 				&	& 2014 May 06 	& GMOS-N		& Blue	& 1$\times$900 &  1.07$-$1.08\\
2MASS J17534518$-$6559559	& L4	& $>50$			&$14.10\pm0.03$	& 2013 Aug 30 	& GMOS-S		& Blue	& 6$\times$900 & 1.23$-$1.25 \\
							& 	& 				&	& 2013 Aug 31 	& GMOS-S		& Blue	& 12$\times$900 & 1.23$-$1.45 \\
2MASS J18212815$+$1414010	& L4.5& $4.2\pm0.1$	&$13.43\pm0.02$	& 2013 Aug 26 	& GMOS-N		& Blue	& 8$\times$900 & 1.01$-$1.09 \\
							& 	  & 				&	& 2013 Aug 29 	& GMOS-N		& Blue	& 6$\times$900 & 1.04$-$1.23 \\
							& 	  & 				&	& 2013 Aug 30 	& GMOS-N		& Blue	& 2$\times$900 & 1.09$-$1.13 \\
2MASS J21481633$+$4003594	& L6	& $19\pm4$		&$14.15\pm0.03$	& 2013 Aug 08 	& GMOS-N		& Blue	& 8$\times$900 & 1.21$-$1.07 \\
							& 	& 				&	& 2013 Sep 01 	& GMOS-N		& Blue	& 5$\times$900 & 1.21$-$1.10 \\
							& 	& 				&	& 2013 Sep 02 	& GMOS-N		& Blue	& 3$\times$900 & 1.06$-$1.08 \\
							& 	& 				&	& 2013 Sep 29 	& GMOS-N		& Blue	& 5$\times$900 & 1.07$-$1.09 \\
\enddata
\tablenotetext{}{{\bf Notes:} $^{\rm a}$ \,Spectral type and rotation period as tabulated in \citet{2015ApJ...799..154M}. $^{\rm b}$ \,Number of exposures $\times$ integration time in seconds. }
\end{deluxetable*}

\section{Observations}
\label{sec:obs}

We used the two copies of the Gemini Multi-Object Spectrograph \citep[GMOS;][]{2004PASP..116..425H} mounted on the 8 m Gemini North and Gemini South telescopes to collect optical spectra of our targets. Observations were carried out in queue mode between August 2013 and May 2014.

Five of our targets (J0103$+$19, J1416$+$13, J1721$+$33, J1821$+$14, and J2148$+$40) were observed from the North using the R831$_{-}$G5302 grating and the remaining three (J1126$-$50, J1254$-$01, and J1753$-$65) from the South with the R831$_{-}$G5322 grating. At both telescopes we used a slit of 0.75$\arcsec$ (pixel scale of 0.08\arcsec) with a binning of 2 pixels,  which yielded a resolution of $\sim$6 \AA~in our spectra. In each campaign we also used the second order-blocking filter GG455 to avoid contamination of our data with stray light from wavelengths $\le$5000 \AA. For six of our targets we used a central wavelength of 5712 \AA\,(``blue configuration''), and for the other two a central wavelength of 7300 \AA~ (``red configuration''). Data collected with the blue and red configurations cover 4670--6820 \AA~and 6270--8460 \AA, respectively. For each target we collected between 8 and 21 individual spectra, by dithering along the spectroscopic slit, with typical exposure times of 780--900 s. On each night we also recorded spectra of Cu$+$Ar lamps with the same instrumental setup as used for the science targets. No standard spectrophotometric stars were observed in any observing epoch, thus, our final spectra were not calibrated in flux. In columns 5--9 of Table \ref{tab1} we provide for each target: dates of observations, instrument used, instrumental configuration, number of individual spectra collected multiplied by individual exposure time, and the range of airmass covered.

\begin{figure*}[]
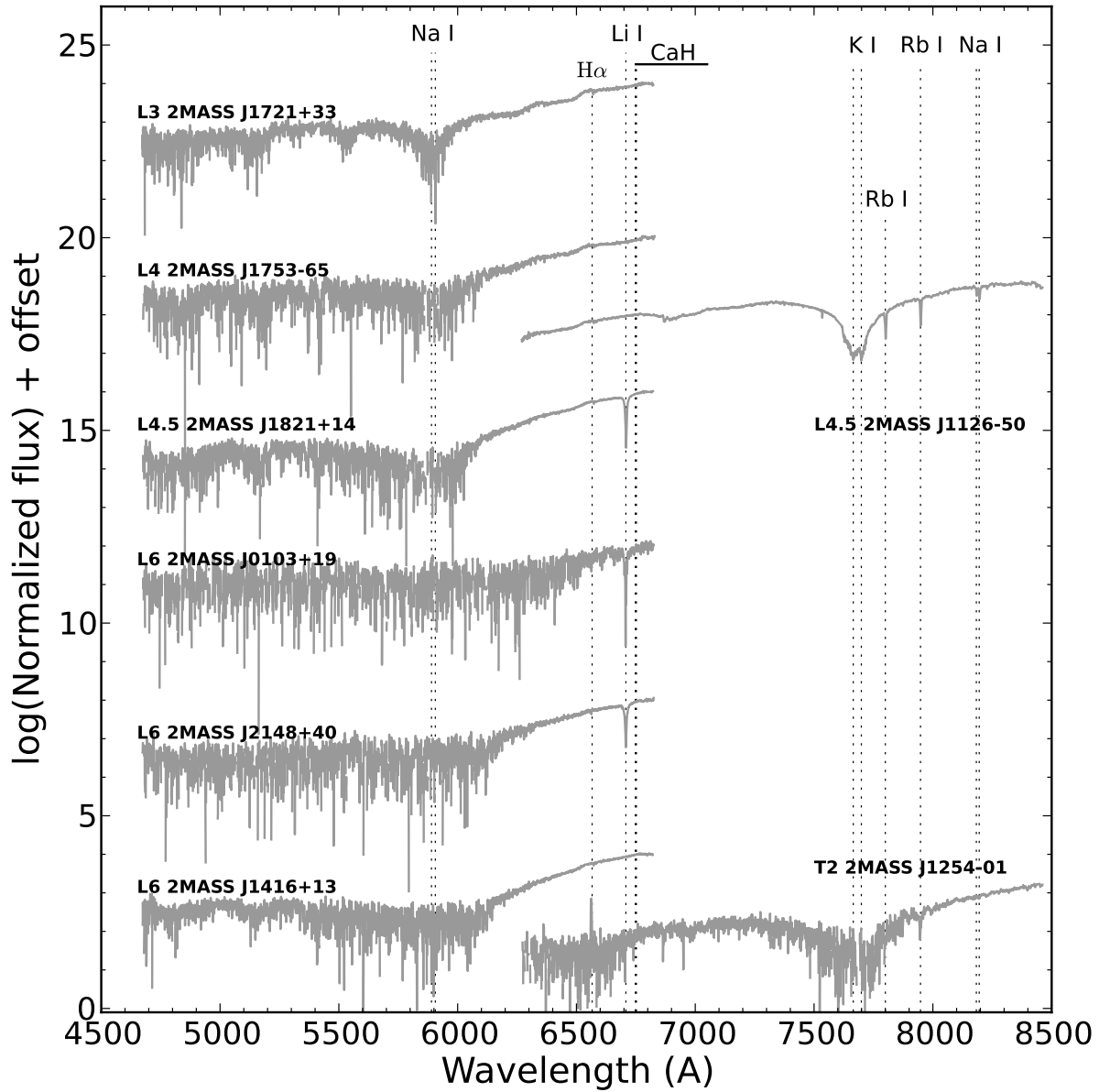

\figurenum{1}
\fig{fig_1_all.eps}{0.9\textwidth}{}
\caption{GMOS spectra of our targets, taken with the blue and red configurations, arranged by spectral type. Spectra were normalized by the average value in the region 6790--6815 \AA. Some atomic lines and molecular features are labeled. \label{fig1}}
\end{figure*}

We used the Gemini package within the Image Reduction and Analysis Facility software (IRAF) for bias subtraction and flat fielding of the data. We extracted the spectra using the IRAF {\sc apextract} standard routines.  We employed typical aperture widths of 20 pixel centered on the spectroscopic traces, which we fit with second- to fourth-order Legendre polynomials.  We estimated sky backgrounds from two 20-pixel-wide bands, centered 20--30 pixels away from the trace of the object.  Figure \ref{fig1} shows the final wavelength-calibrated (with a typical uncertainty of $\pm$0.1--0.2 \AA) and median-combined spectra for all eight targets.

\section{Results from our survey}
\label{sec:observed_sample}

\subsection{H$\alpha$ emission}
\label{sec:dis_halpha}

\begin{figure}[]
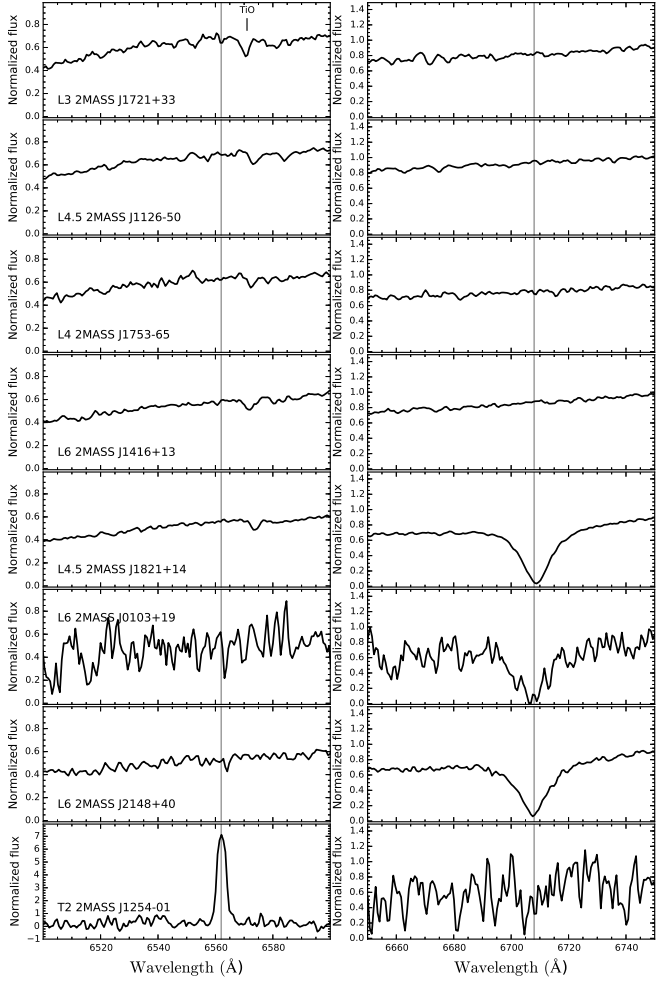

\figurenum{2}
  \fig{fig_2_zoom.eps}{0.49\textwidth}{}
\caption{Normalized spectra of our 8 targets zoomed in the regions that contain the H$\alpha$ (left) and the lithium (right) lines. The vertical scale is the same for all targets with the exception of J1254--01.\label{fig2}}
\end{figure}

The main motivation of our survey is the search for H$\alpha$ emission in a set of targets for which we have exquisitely precise determinations of the presence or absence of photometric variability.  As seen in Figure \ref{fig1}, we detect H$\alpha$ in only one of the eight targets: the field T2 dwarf J1254$-$01, which is not known to vary photometrically.  H$\alpha$ emission from this T2 dwarf was already reported by  \citet{2003ApJ...594..510B}.   The remaining seven targets in our sample---all L dwarfs---do not show any sign of H$\alpha$ emission, even though five of them exhibit photometric variability compatible with rotation.  

Figure \ref{fig2} shows the locations of the H$\alpha$ $\lambda6563$\,\AA\, and \ion{Li}{1} $\lambda6708$\,\AA\, lines.  H$\alpha$ emission is not detectable even at low levels among the seven L dwarfs.  We collected spectra of most of our targets, with the exception of the L6 dwarf J1416$+$13, on at least two different nights.  Weak signs of H$\alpha$ are not seen in the individual spectra, either. Therefore, it is unlikely that we have missed a phase of emission. Note that the photospheric H$\alpha$ absorption---seen from F to mid M stars---is not expected in the spectra of ultra-cool dwarfs ($\ge$M7) since their photospheres are too cool to populate the $n=2$ level of hydrogen significantly. Thus, H$\alpha$ can only be present as an emission line if there is a chromospheric heating \citep{1985ApJ...294..626C,1985ApJ...299..781G,1986ApJ...305..784G,2000ARA&A..38..485B}.

We measured the pseudo-equivalent width (pEW) of the H$\alpha$ emission line detected in J1254--01 by first  subtracting the neighboring continuum determined from a 45 \AA-wide region (excluding the 10 \AA\, centered in the H$\alpha$ line), and then summing the flux contained in the central 10 \AA\, as done in \citet{2016ApJ...826...73P}. We also integrated the flux in different regions of the continuum and adopted the average value as the uncertainty in our pEW. For the remaining 7 targets without H$\alpha$ emission, we adopted the same procedure for estimating 3$\sigma$ upper limits. These values are given in Table \ref{ew}.

\begin{table}[h!]
\renewcommand{\thetable}{\arabic{table}}
\centering
\caption{Pseudo-equivalent widths and 3$\sigma$ upper limits for H$\alpha$ and \ion{Li}{1}\label{ew}}
\begin{tabular}{lcc}
\tablewidth{0pt}
\hline
\hline
Object & H$\alpha$ & \ion{Li}{1}   \\
           & (\AA)&  (\AA)  \\
\hline
J0103$+$1935 &$>\,$-0.6 &12.9$\,\pm\,$0.2 \\
J1126$-$5003 &$>\,$-0.3 & $\le\,$0.3\\
J1254$+$0122 &-26.6$\,\pm\,$0.2 & $\le\,$0.5 \\
J1416$+$1348 &$>\,$-0.3 & $\le\,$0.3\\
J1721$+$3344 &$>\,$-0.3 & $\le\,$0.3\\
J1753$-$6559 &$>\,$-0.3 & $\le\,$0.3\\
J1821$+$1414 &$>\,$-0.3 & 13.5$\,\pm\,$0.1 \\
J2148$+$4003 &$>\,$-0.3 & 12.6$\,\pm\,$0.1 \\
\decimals
\hline 
\end{tabular}
\\
\raggedright
\scriptsize
\end{table}

\begin{figure}[h]
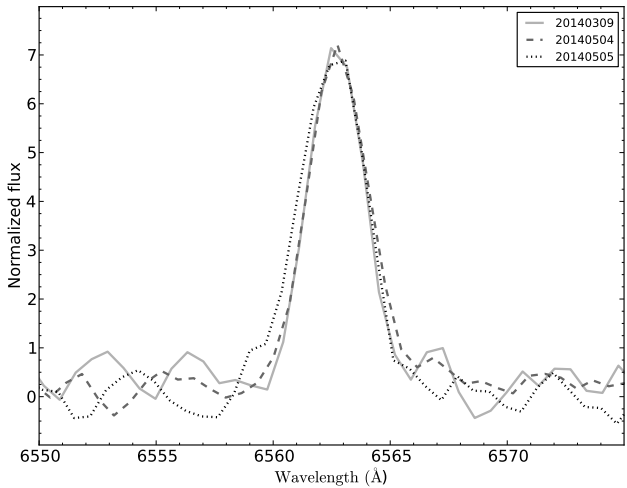

\figurenum{3}
  \fig{fig_j1254.eps}{0.49\textwidth}{}
\caption{Normalized spectra of J1254--01 zoomed in the H$\alpha$ line, taken in 2014 (March 9: black; May 4: blue; and May 5: green). The H$\alpha$ emission does not show any significant variation in the $\sim$2 months that separate the first measurement from the second and third ones.\label{fig3}}
\end{figure}

The one target from which we detect H$\alpha$ emission, the non-photometrically variable T2 dwarf J1254$-$01, was observed at three different epochs: one in March 2014, and two on consecutive days in May 2014 (Table \ref{tab1}). We do not detect any obvious changes in the H$\alpha$ strength during the $\sim$2 months that separate the first and the last measurements (Figure~\ref{fig3}). This may point to the presence of a stable emission region on this brown dwarf. To investigate the importance of the chromospheric phenomena in the total energy budget of J1254--01 we computed its H$\alpha$ to bolometric flux ratio. We measured a value of $(4.3\pm2.0)\times10^{-19}$ erg\,cm$^{-2}$\,s$^{-1}$\,\AA$^{-1}$ for the surrounding continuum of H$\alpha$ in the flux-calibrated spectra of J1254--01 presented in \citet{2003ApJ...594..510B}, which in combination with our pEW yielded F$_\alpha\,=(11.6\pm5.3)\times10^{-18}$ erg cm$^{-2}$ s$^{-1}$. Then we used the distance and bolometric luminosity given in  \citet{2015ApJ...810..158F} to obtain log$(L_\alpha/L_{\rm bol})=-5.6\pm0.2$. Previous works found log$(L_\alpha/L_{\rm bol})$ values of --5.8 \citep[][]{2003ApJ...594..510B} and $-5.9\pm0.2$ \citep[][]{2016ApJ...826...73P}, which are in agreement with our determination and suggest that the emission region of J1254--01 could have remained stable since the first H$\alpha$ observations by \citet[][]{2003ApJ...594..510B} 13.3 years ago. 
 
Interestingly, J1254--01 did not show any photometric variability {within} $\pm$0.15\,\% (3.6 $\mu$m) or $\pm$0.3\,\% (4.5 $\mu$m) during 19 h of continuous monitoring with {\sl Spitzer} \citep{2015ApJ...799..154M}, or {within} $\pm$1.8\,\% ($I$ band) or $\pm$4.8\,\% ($Z$ band)  in two observing runs of 4 h each \citep{2015ApJ...801..104H}. \citet{2013ApJ...767...77S} compared the infrared absolute flux  of this dwarf to models, and estimated a radius of $0.84\,\pm\,0.05$ R$_{\rm J}$, which indicates a high surface gravity \citep[log\,{\sl g}$\,\ge$5.0,][]{2004AJ....127.3516G,2008ApJ...678.1372C} and hence, an old object.  Combined with the lack of H$\alpha$ variability in our observations, accretion of circumstellar material is unlikely as an origin of the H$\alpha$ emission.   \citet{2006ApJ...647.1405Z} reported a projected rotational velocity $v\sin i = 28.4\pm2.8$ km s$^{-1}$.  Combined with the radius estimate, this places an upper limit on the rotation period of 3.6 h. This could have been detected by \cite{2015ApJ...799..154M} or \cite{2015ApJ...801..104H} unless the object were not variable at the time of the observations, or unless its rotation axis is strongly inclined so as to hinder the detection of photometric modulations caused by spots that are either always or never visible.  Regardless, the combination of strong but unvarying H$\alpha$ emission and the lack of optical and infrared photometric variability at multiple epochs suggests that the H$\alpha$ emission mechanism is unrelated to rotationally modulated variability.

\subsection{Neutral lithium absorption}
\label{sec:dis_lithium}

Three of our targets (J0103$+$19, J1821$+$14, and J2148$+$40) show strong lithium absorption at 6708 \AA, as already reported in their discovery papers \citep{2000AJ....120..447K,2008ApJ...686..528L}. This confirms their substellar  nature and masses $\le\,$0.060 M$_{\sun}$ independently of their age \citep[][]{1992ApJ...389L..83R,2000ARA&A..38..485B,2008ApJ...689.1295K}. We measured their pEWs by integrating the fluxes of the objects between 6695 \AA\, and 6722 \AA. The associated uncertainties and 3$\sigma$ upper limits for the remaining 5 targets with no obvious absorption were computed as for H$\alpha$ emission in Section~\ref{sec:dis_halpha}. These values are also listed in Table \ref{ew}. Our pEW measurements are in good agreement with those presented in the corresponding discovery papers of J0103$+$19 \citep[12 \AA, ][]{2000AJ....120..447K}, J1821$+$14  \citep[$13.9\pm0.4$ \AA, ][]{2008ApJ...686..528L}, and J2148$+$40 \citep[$12.1\pm0.6$ \AA, ][]{2008ApJ...686..528L}.

\section{Enlarging the sample with previous studies}
\label{sec:literature_sample}
The lack of H$\alpha$ emission in our six variable targets suggests that there is no evident correlation between the existence of magnetic activity and photometric variability. However, a larger sample is needed to attain more robust conclusions. We searched the literature for all L and T dwarfs with measurements of both optical spectra containing the H$\alpha$ region and optical or near-infrared photometric monitoring lasting at least 2 h (comparable to the shortest rotation periods observed in ultra-cool dwarfs; \citealt{2015ApJ...799..154M}) and photometric accuracy $\le$5 \%. We found 66 L and 20 T dwarfs that fulfill these requirements. Combined with our eight targets, these result in 73 L and 21 T dwarfs.  Table \ref{tab2} lists their full identifiers, spectral types, indications of detection$/$non-detection of H$\alpha$ and photometric variability, rotation periods (if measured), filters for the photometric monitoring, duration of monitoring, photometric accuracy, and  their references.

\begin{table}[h!]
\renewcommand{\thetable}{\arabic{table}}
\centering
\caption{Detections and non-detections of H$\alpha$ or photometric variability in our combined sample of observed targets and literature targets.\label{tab3}}
\begin{tabular}{lccccc}
\tablewidth{0pt}
\hline
\hline
SpT range & Objects & YY$^{\rm a}$ & YN$^{\rm b}$  & NY$^{\rm c}$  & NN$^{\rm d}$ \\
\hline
\decimals
L0--L1.5	&25	&8	&17	&0	&0 \\
L2--L3.5	&24	&6	&15	&2	&1 \\
L4--L5.5 &14 	&1 	&4 	&6 	&3 \\
L6--L9.5	&10 	&0 	&1	&5	&4 \\
T0--T4 	&11 	&0	&1	&6	&4 \\
T4.5--T8	&10	&1	&1	&2	&6 \\
\hline 
\end{tabular}
\\
\raggedright
\scriptsize
{\bf Notes:} $^{\rm a}$ Objects with both H$\alpha$ emission and photometric variability. \\
$^{\rm b}$ Objects with H$\alpha$ emission but without photometric variability. \\
$^{\rm c}$ Objects without H$\alpha$ emission, but with photometric variability; \\
$^{\rm d}$ Objects displaying neither H$\alpha$ emission  nor photometric variability.
\end{table}

We visually inspected the light curves of each of these objects to check that they display a convincing variable or non-variable (``flat'') light curve within the photometric error as claimed in their references, and similarly for the detections of H$\alpha$ emission. Given the large number of groups that have contributed to the information listed in Table \ref{tab2}, the range of spectroscopic and photometric accuracies is wide. The homogenization of these ranges is not a trivial task, and we did not attempt to apply any sensitivity correction. The following discussion is based simply on the number of objects with reported detections or non-detections of either H$\alpha$ emission or photometric variability. All of the statistics that are discussed below are also synthesized in Table \ref{tab3}, where we provide the combined number of objects from our survey and from the literature in a series of spectral type bins: L0--L1.5, L2--L3.5, L4--L5.5, L6--L9.5, T0--T4, and T4.5--T8. For each of these bins Table~\ref{tab3} lists the number of objects that are reported to show H$\alpha$ emission with or without photometric variability, and the number of objects that are reported to photometrically vary with or without H$\alpha$ emission.

\section{Discussion of H$\alpha$ emission and photometric variability}
\label{sec:discussion}

Armed with the expanded statistics of the enlarged sample, we address two related questions: (1) whether a single mechanism can explain both H$\alpha$ emission and photometric variability across all L and T dwarfs (Section~\ref{sec:dis_nosingle}), and (2) whether H$\alpha$ emission may still be correlated with the detection of photometric variability in early L dwarfs (Section~\ref{sec:dis_noenhanced}).  We then discuss how the H$\alpha$ emission and the photometric variability mechanisms may differ (Section~\ref{sec:dis_mechanisms}).

\subsection{No single mechanism for H$\alpha$ emission and photometric variability across L and T dwarfs}
\label{sec:dis_nosingle}

\begin{figure}[]
\figurenum{4}
\gridline{\fig{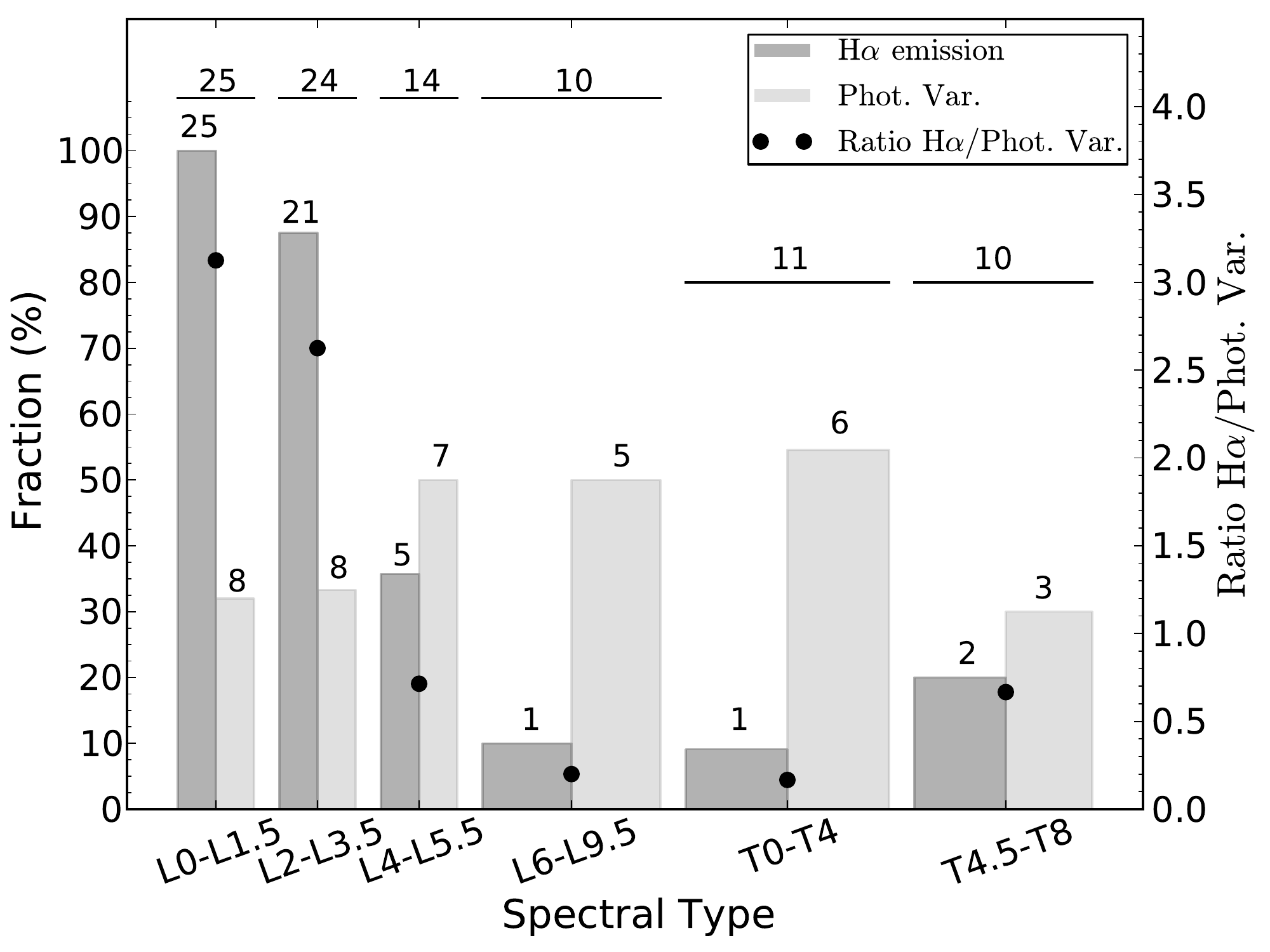}{0.49\textwidth}{}}
\gridline{\fig{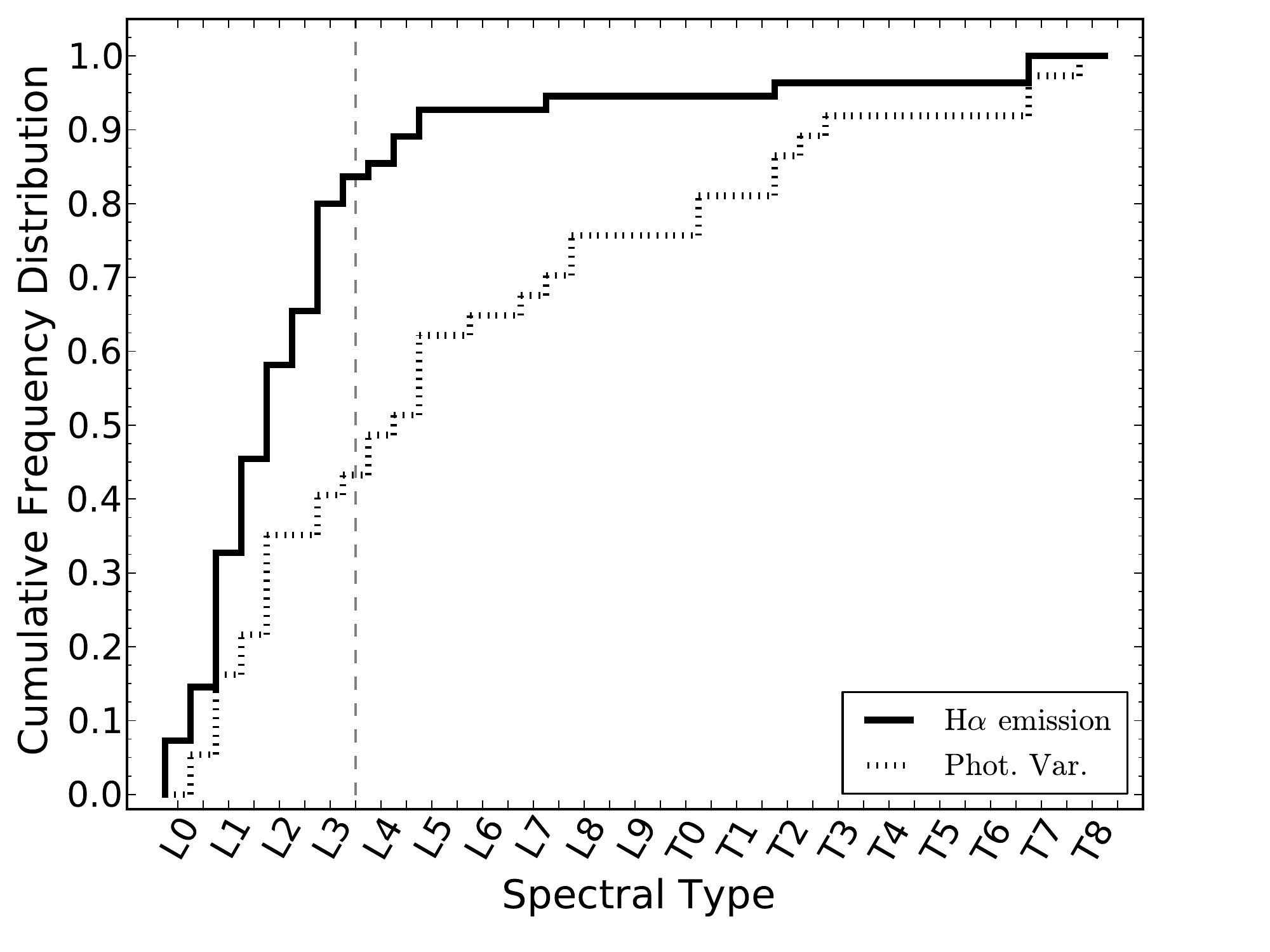}{0.49\textwidth}{}}        
\caption{{\sl Top:} Detection statistics of H$\alpha$ emission (dark gray) and photometric variability (light gray) as a function of spectral type.  The numbers of objects in each bin are given above the horizontal lines at the top.  The numbers of objects with H$\alpha$ emission or photometric variability are given above their respective histogram bars. Black dots denote the ratios of objects with H$\alpha$ emission over objects with photometric variability per bin ($y$-axis on the right). {\sl Bottom:} Cumulative frequency distributions (CFD) of the H$\alpha$ emission and photometric variability statistics. The greatest difference between the two CFDs occurs at spectral type L3.5 (shown with a vertical dashed line). \label{fig5}}
\end{figure}

The top panel of Figure \ref{fig5} shows the detection rates of H$\alpha$ emission and photometric variability per spectral bin (Table \ref{tab3}); the ratio of these two rates is also shown.  The H$\alpha$ emission rate is close to 100\% at L0--L3.5 types, and drops to $\leq$35\% at spectral types $\geq$L4.   In contrast, the rate of detected photometric variability is approximately unchanged, between 30--50\% across all spectral types as already reported by \citet{2014ApJ...782...77B} and \citet{2015ApJ...799..154M}.  

We use the Kolmogorov-Smirnov (K-S) two-sample test \citep{2003psa..book.....W} to test if these observables arise from a common distribution function or are statistically different. The cumulative frequency distributions (CFDs) for the H$\alpha$ and photometric variability detections are shown in the bottom panel of Figure \ref{fig5}.  The K-S test rejects the null hypothesis---that both observables are drawn from the same parent distribution---at the 99.9 \% level.  We conclude that a single mechanism can not explain the presence of both H$\alpha$ emission and photometric variability in L0--T8 dwarfs.

\subsection{No enhanced photometric variability in H$\alpha$-emitting early-L dwarfs}
\label{sec:dis_noenhanced}

The bottom panel of Figure~\ref{fig5} affirms that the maximum change in the relative frequencies of H$\alpha$ emission and photometric variability occurs at spectral type L3.5. We test whether the proportions of H$\alpha$ emission for L0--L3.5 dwarfs ($p_1$) and L4--T8 dwarfs ($p_2$) are statistically different by using a classical two-tailed hypothesis test \citep{2012msma.book.....F}, which assumes $p_1=p_2$ as the null hypothesis, and $p_1\neq\,p_2$ as the alternative. For this test, the null hypothesis is rejected with a confidence of 99.9\,$\%$ if the test statistic, $z$, satisfies:  $z\,=\,(p_{1}-p_{2})/\sqrt{\frac{p_{1}\times(1-p_{1})}{n_1}+\frac{p_{2}\times(1-p_{2})}{n_2}}\ge3.29$, where $n_1$ and $n_2$ are the total number of ultra-cool dwarfs in the L0--L3.5 and L4--T8 spectral ranges, respectively. From Table \ref{tab3}, $p_1 =  46/49$ and $p_2 = 9/45$, which results in $z=10.7$. Thus, we can reject the null hypothesis that the rate of H$\alpha$ detections is the same in the spectral ranges L0--L3.5 and L4--T8  with a confidence of $>$99.9 $\%$. 

The drop in the detection rate of H$\alpha$ activity at $\geq$L4 spectral types has also been reported by \citet{2015AJ....149..158S}.  These authors note that the activity fraction increases from early M dwarfs peaking at L0---where 90 \% of dwarfs seem to be active---and then drops to being negligible, or below their detection limits, for L4--L8 dwarfs. It may be argued that cooler brown dwarfs are intrinsically fainter, and hence H$\alpha$ emission may be more difficult to detect.  However, the same argument should then hold for the detectability of photometric variations.  We note that similar proportions of H$\alpha$-emitting ultra-cool dwarfs lack detected photometric variability in the L0--L3.5 (32 out of 46) and in the L4--T8 (7 out of 9) ranges.\footnote{These proportions correspond to the YN/(YY + YN) ratio formed from Table~\ref{tab3}, summed over the two spectral type ranges.}  These proportions would be expected to be different if SNR played a role.  If target brightness mattered, we should be seeing higher incidence of photometric variability among targets with H$\alpha$ detections. 

We conclude that the elevated H$\alpha$ activity among L0--L3.5 dwarfs is not linked to a higher incidence of photometric variability. This conclusion is based on the statistics of our enlarged sample, but it is also supported by observations and radiative transfer models (Section \ref{sec:intro}) that probe that photometric variability cannot be explained solely by localized heating from reconnection events.

\subsection{Disentangling the H$\alpha$ emission and photometric variability mechanisms}
\label{sec:dis_mechanisms}

The conclusions from Sections~\ref{sec:dis_nosingle} and \ref{sec:dis_noenhanced} indicate that H$\alpha$ emission and photometric variability in L--T dwarfs are driven by distinct mechanisms.  

A likely explanation for the decrease in H$\alpha$ activity toward later L types is the strong increase of atmospheric resistivity at effective temperatures $<$2300 K \citep[i.e. spectral types cooler than M9$/$L0,][]{2002ApJ...571..469M}.  As the number of collisions between neutral and charged particles (from atomic ionization) increases with dropping effective temperature, the collisions hinder the generation and transport of currents through the atmosphere.  The result is a reduction in the energy available to support a chromosphere and in the generation of H$\alpha$ emission. For spectral types $\ge$L4 the resistivity is so high that activity drops severely. The sporadic detections of activity in this range could be the result of the interaction of dust particles that partially ionize the surrounding medium, leading to transient chromospheric features and H$\alpha$ emission \citep{2011ApJ...737...38H,2013ApJ...767..136H}. 

As much as magnetic activity is an attractive option to account for photometric variability in H$\alpha$-emitting objects, especially among early-L dwarfs, the ionization levels of these atmospheres are so low that it is very unlikely that the magnetic field is coupled to the atmosphere. Thus, magnetic fields will not either drive cloud formation or drag structures in the atmosphere with it, which can explain why the detection rate of photometric variability is similar across the L0--T8 spectral range, and not larger for the earliest L dwarfs given their higher activity rate. Moreover, observations at optical and near-infrared wavelengths probe low-pressure regions in the atmosphere typically from $\sim$0.1 to 10 bar \citep{2001ApJ...556..357A,2011ApJ...737...34M,2016ApJ...826....8Y}, in which the radiative timescales are expected to be short: a thermal incursion introduced by reconnection events is expected to decay in a $\sim$1 h timescale as seen, for example, in the flares presented by \cite{2013ApJ...779..172G} and \cite{2015AJ....149..104B}. Because of this, localized heating cannot be responsible for the structures that are seen over significantly longer timescales on many ultra-cool dwarfs.

Conversely, given the similar photometric variability detection rates from L0 to T8 dwarfs, the dominant mechanism for producing large-amplitude photometric variations may be the same throughout: dust clouds. In fact, there is evidence that dust clouds---predicted to be present in dwarfs $\ge$M7 \citep{1996AA...308L..29T}---are important in late-M dwarfs and early-L dwarfs. For example, the M9 dwarf TVLM~513--46546 harbors a multipolar magnetic field with intensities as high as 3 kG \citep{2008ApJ...673.1080B}, and displays both an optical light curve, that has remained stable for at least 5 yr \citep{2013ApJ...779..101H}, and radio emission \citep{2007ApJ...663L..25H}.  While these characteristics hint at a role for the magnetic field in setting the inhomogeneities by means, for example, of optical and radio aurorae \citep{2015Natur.523..568H,2016ApJ...818...24K}, \citet{2008MNRAS.391L..88L} showed that the dwarf's Sloan-$g$ and Sloan-$i$ light curves are anti-correlated: a fact that is incompatible with a star spot scenario and is better explained with the existence of a dust cloud. In addition, \citet{2015A&A...580L..12M} found that the $I$ band linear polarimetric light curve of TVLM~513--46546 changes from $\le$0.3 \% to $\sim$1.3 \%, with the same periodicity as the photometric variations, which is better explained by a dust cloud located at the photosphere. Other two examples of early L dwarfs displaying both magnetic activity and photometric variability \citep{2013ApJ...767..173H,2015ApJ...813..104G} were presented in Section \ref{sec:intro}. In both cases, the best explanation for the observed variability invokes the existence of a dust cloud. Similarly for cooler spectral types, \citet{2013ApJ...768..121A} showed that the variations of the {\sl Hubble Space Telescope} near-infrared spectra of two L/T dwarfs are better explained by the combination of variations of both temperature and clouds.

\section{Conclusions}
\label{sec:conclusion}

We searched for H$\alpha$ emission in a sample of eight L3--T2 dwarfs, all of which had been monitored to a high precision photometrically for 16--21 h in the  [3.6] and [4.5] {\sl Spitzer} bands, to seek a correlation between magnetic activity and photometric variability. We detected H$\alpha$ emission only in the T2 dwarf  J1254--01, which has not been reported as a photometric variable in the literature at either optical or infrared wavelengths. Its H$\alpha$ emission feature is constant within our observational accuracy for $\sim$2 months, and there is a hint that it could have remained nearly unchanged for at least 13.3 years.  Among the remaining seven L dwarfs of our survey, six are known to vary photometrically, yet did not display any H$\alpha$ emission.

We expanded our initial sample with data from the literature, yielding a total sample of 94 L and T dwarfs with both spectroscopic measurements containing the H$\alpha$ region and with photometric monitoring for at least 2 h at optical and infrared wavelengths. We found that the observed rates of H$\alpha$ emission and photometric variability follow significantly different dependencies as a function of spectral type.  The H$\alpha$ detection rate drops sharply from 88\%--100\% at $\leq$L3.5 to $<$35\% at $\geq$L4, while the detection rate of photometric variability is roughly uniform (30\%--55\%) over L0--T8. The disparate behavior of H$\alpha$ emission and photometric variability---both in our high-sensitivity mini-sample of eight L3--T2 dwarfs and in the broader meta-sample of 94 L0--T8 dwarfs---indicates that these two phenomena are uncorrelated.  Consequently, even early-L dwarfs, among which H$\alpha$ emission is nearly ubiquitous, likely have their largest photometric variations driven by dust clouds.  These conclusions are supported by previous multi-wavelength analyses of variable M9--L3 dwarfs, which disfavor hot chromospheric spots as drivers for the photometric variability.

\acknowledgments

We are thankful to the anonymous referee for his/her valuable comments. We are also grateful to the Gemini Observing Assistants, Instrument Specialists and Support Astronomers that performed the data acquisition of our targets. IRAF is distributed by the National Optical Astronomy Observatories, which are operated by the Association of Universities for Research in Astronomy, Inc., under cooperative agreement with the National Science Foundation. This work is based in part on observations made with the {\sl Spitzer Space Telescope}, which is operated by the Jet Propulsion Laboratory, California Institute of Technology under a contract with NASA. Support for this work was provided by NASA through awards issued by JPL/Caltech.

%\floattable
\begin{deluxetable*}{lcccccccc}
\tablecaption{H$\alpha$ and photometric variability detections and non-detections in the literature. \label{tab2}}
\tablecolumns{9}
\tablenum{2}
\tablewidth{0pt}
\colnumbers
\tabletypesize{\scriptsize}
\tablehead{
\colhead{Object} &
\colhead{SpT} &
\colhead{H$\alpha$?} &
\colhead{Phot. var.?} &
\colhead{Rot. Per. (h)} &
\colhead{Filter} &
\colhead{Duration (h)} &
\colhead{$\sigma_{\rm phot}$ ($\%$)} &
\colhead{Ref.} 
}
\startdata
2MASS J00584253$-$0651239    &    L0    &    Y    &    N    & \nodata  &  $I$ Cousins & 2.6 & 1.2  &3, 29 \\       
2MASS J03140344$+$1603056    &    L0    &    Y    &    N    & \nodata & $I$ Cousins & 2.0 & 0.7  & 21, 29 \\       
2MASS J11593850$+$0057268    &    L0    &    Y    &    N    & \nodata  & $I$ Cousins & 3.0 & 0.6  & 21, 29 \\       
2MASS J22000201$-$3038327    &    L0    &    Y    &    N    & \nodata &   $I$ Cousins & 3.2 & 4.4  &21, 29 \\       
2MASS J07464256$+$2000321A    &    L0.5    &    Y    &    Y    &    3.32$\pm$0.15   & $I$ & 61.6\tablenotemark{a} & 0.3$^{\rm a}$  & 21, 30 \\
2MASS J14122449$+$1633115    &    L0.5    &    Y    &    N    & \nodata  &  $I$ Cousins & 3.3 & 1.6  &21, 7 \\       
2MASS J14413716$-$0945590    &    L0.5    &    Y    &    N    &  \nodata  &  $I$ Cousins  & 8.2 & 1.2  & 28, 29 \\   
2MASS J23515044$-$2537367B    &    L0.5    &    Y    &    Y    &  \nodata & $I$ Cousins  & 7.0 & 0.9  & 21, 29 \\   
2MASS J10224821$+$5825453    &    L1    &    Y    &    Y    & \nodata  & Sloan $i$ & 3.5 & 1.4  & 21, 38 \\    
2MASS J1045240$-$014957    &    L1    &    Y    &    N    &  \nodata  &  $I$ Cousins  & 6.7 & 0.6  &28, 29 \\   
2MASS J10484281$+$0111580     &    L1    &    Y    &    Y    &  \nodata & $I$ Cousins  & 2.6 &  1.0 &  21, 29 \\   
2MASSW J1108307$+$683017    &    L1    &    Y    &    N    &  \nodata &   Sloan $i$ & 4.5 & 1.0  &4, 38 \\   
2MASS J14392836$+$1929149    &    L1    &    Y    &    N    &  \nodata &  $R$, $I$ Cousins  & 6.8, 1.2  & 1.1, 0.9   & 5, 9, 15, 21, 29 \\   
2MASS J15551573$-$0956055    &    L1    &    Y    &    N    &  \nodata &  $I$ Cousins  & 11.1$^{\rm b}$ & 0.5  & 16, 21 \\   
WISEP J190648.47$+$401106.8    &    L1    &    Y     &    Y    &    8.9 h &   {\sl Kepler} & 2.3 yr & 0.5  &40 \\
2MASS J1300425 $+$ 191235    &    L1    &    Y    &    Y    &  \nodata &   R, $I$ Cousins$^{\rm c}$ & 5.6, 22.9 & 1.1, 1.4   &7, 15, 17, 28, 33 \\   
2MASSI J0829066$+$145622    &    L1    &    Y    &    N    &  \nodata &  $I$ Cousins  & 3.2 & 0.8  &29, 36 \\   
DENIS J090957.1$-$065806    &    L1    &    Y    &    N    & \nodata &   $I$ Cousins & 6.3 & 0.9  &8, 29, 41 \\    
2MASS J11455714$+$2317297    &    L1.5    &    Y    &    Y    &   \nodata & $I$ Cousins & 9.3 & 1.8  &  5, 21 \\  
2MASS J1334062$+$194035    &    L1.5    &    Y    &    Y    &   \nodata &    $I$ Cousins & 6.8 & 1.8  &5, 28\\  
2MASS J16452211$-$1319516    &    L1.5    &    Y    &    N    &    \nodata & $I$ Cousins & 3.1 & 1.0  & 11, 21 \\ 
2MASS J20575409$-$0252302    &    L1.5    &    Y    &    N    &    \nodata & $I$ Cousins & 4.4 & 1.0  &  11, 17, 21 \\ 
2MASSW J0832045$-$012835    &    L1.5    &    Y    &    N    &    \nodata &   $I$ Cousins & 2.0 & 1.1  &3, 29 \\ 
2MASSW J0135358$+$120522    &    L1.5    &    Y    &    N    &    \nodata &  Sloan $i$& 2.2 & 1.8  &3, 38 \\ 
DENIS J174534.6$-$164053    &    L1.5    &    Y    &    N    &    \nodata &  $I$ Cousins & 2.7 &  1.4 & 29, 41 \\ 
2MASS J08283419$-$1309198    &    L2    &    Y     &    Y    &    2.9 &  $I$ Cousins  & 11.4$^{\rm d}$ & 1.3  &14, 21 \\
2MASS J0921141$-$210444    &    L2    &    Y    &    Y    &    \nodata &  $R$, $I$ Cousins & 1.4, 1.4 & 1.8, 1.0  & 28, 29 \\ 
2MASS J11553952$-$3727350    &    L2    &    Y    &    Y    &    $\sim$8 & $I$ Cousins  & 23.8$^{\rm e}$ & 0.5  &11, 21 \\
Kelu$-$1    &    L2    &    Y    &    Y    &    1.8$\pm$0.05 &  $I$ & 13.0$^{\rm f}$ & 0.1  & 8, 12, 21 \\
2MASS J00154476$+$3516026    &    L2    &    Y    &    N    &    \nodata &  Sloan $i$ & 2.8 & 0.5  & 3, 38 \\ 
2MASSW J0030438$+$313932    &    L2    &    Y    &    N    &    \nodata &   $I$ & 4.9 & 2.3  &1, 5 \\ 
2MASSI J0847287$-$153237    &    L2    &    Y    &    N    &    \nodata &  $I$ Cousins & 7.6 & 1.0  & 29, 41 \\ 
2MASSI J1726000$+$153819    &    L2    &    N    &    N    &    \nodata  & [3.6], [4.5] & 14, 7 & 0.29, 0.49  & 3, 37 \\ 
2MASSW J2208136$+$292121    &    L2    &    N    &    Y    &    3.5$\pm$0.2  &  [3.6], [4.5] & 14, 7 & 0.07, 0.11  & 3, 37 \\
2MASS J05233822$-$1403022    &    L2.5    &    Y    &    N    &    \nodata &  $I$ Cousins & 7.0 &  0.8 & 21, 29, 33 \\ 
2MASS J10292165$+$1626526    &    L2.5    &    Y    &    N    &    \nodata & $R$, $I$ & 2.5, 2.2 & 1.8, 1.1  & 19, 21 \\ 
2MASS J1047310$−$181557    &    L2.5    &    Y    &    N    &    \nodata &  $R$, $I$ Cousins & 3.5, 13.6$^{\rm g}$ & 2.3, 1.0  & 28, 29 \\ 
DENIS J081231.6$-$244442    &    L2.5    &    Y    &    N    &    \nodata &  $R$, $I$ Cousins & 3.4, 5.7 & 2.5, 0.9  & 41, 29 \\ 
2MASS J10584787$-$1548172    &    L3    &    Y    &    Y    &    4.1$\pm$0.2 &  [3.6], [4.5] & 8, 6 & 0.2, 0.2  & 1, 37 \\
2MASS J0913032$+$184150    &    L3     &    Y   &    N    &    \nodata &  $I$ & 5.4 & 3.5  & 5, 28 \\ 
2MASS J1203581$+$001550    &    L3     &    Y    &    N    &    \nodata & $I$ & 6.8 & 0.6  &  5, 28 \\ 
2MASS J15065441$+$1321060    &    L3     &    Y    &    N    &    \nodata & $R$, $I$ & 4.5, 5.2  & 1.8, 0.9  &  4, 19, 21 \\ 
2MASS J1615441$+$355900    &    L3     &    Y    &    N    &    \nodata &  $I$ Cousins & 3.6 & 6.0  & 7, 28 \\ 
2MASS J2104149$−$103736    &    L3     &    Y    &    N    &    \nodata & $I$ Cousins & 16.4$^{\rm h}$ & 1.1  &  28, 29 \\ 
2MASS J08355829$+$0548308    &    L3    &    Y    &    N    &    \nodata &  $I$ Cousins & 6.5 & 1.5  &29, 36 \\ 
2MASS J11463449$+$2230527    &    L3    &    Y    &    N    &    \nodata &  $I$ & 3.9 & 1.0  & 5, 36 \\ 
2MASS J03261367$+$2950152    &    L3.5    &    Y    &    N    &    \nodata &  $I$ & 2.5 & 1.6  & 1, 5 \\ 
2MASS J00361617$+$1821104    &    L3.5    &    Y    &    Y    &    2.7$\pm$0.3 & $I$, [3.6], [4.5]  & 10.5, 8, 6 & 0.9, 0.1, 0.1 & 30, 37, 43 \\
2MASS J1705483$−$051646    &    L4    &    Y    &    N    &    \nodata &   $J$ & 3.04 &  0.02 &28, 45  \\ 
2MASS J161542552$+$49532117    &    L4    &    N    &    Y    &    24 &   $I$ & 3.6 & 5.9  &7, 28 \\
DENIS 1228$-$1547    &    L4.5    &    Y    &    N    &    \nodata &   $I$ Cousins, $J$ & 4.3, 3.4 & 1.3, 0.7  & 12, 29, 33, 44 \\ 
2MASSW J2224438$-$015852    &    L4.5    &    Y    &    N    &   \nodata & [3.6], [4.5] & 12, 8 & 0.14, 0.14  &  3, 37 \\ 
2MASS J13153094−2649513    &    L5    &    Y    &    N    &    \nodata &  $I$ Cousins & 2.9 & 1.3  & 27, 29 \\ 
2MASS J01443536$-$0716142    &    L5    &    Y    &    Y    &    aperiodic &  $I$ Cousins & 14.5$^{\rm k}$ & 1.5  & 13, 29 \\
2MASSI J0421072$-$630602    &    L5    &    N    &    N    &    \nodata &   [3.6], [4.5] & 14, 7 & 0.2, 0.2  & 24, 37 \\ 
2MASSW J0820299$+$450031    &    L5    &    N    &    N    &    \nodata & [3.6], [4.5] & 14, 7 & 0.4, 0.4  &  3, 37 \\ 
DENIS$-$P J142527.97$-$365023.4    &    L5    &    N    &    Y    &    3.7$\pm$0.8 & $J$ & 3.24 & 0.02  & 18, 45 \\
2MASSW J1507476$-$162738    &    L5    &    N    &    Y    &    2.5$\pm$0.1 & [3.6], [4.5] & 12, 8 & 0.14, 0.14  & 3, 22, 28, 37 \\
2MASS J11501322$+$0520124    &    L6    &    N    &    N    &    \nodata & [3.6], [4.5] & 14, 7 & 0.5, 0.6  &  26, 37 \\ 
2MASSI J0825196$+$211552    &    L7.5    &    N    &    Y    &    7.6 &  [3.6], [4.5] & 12, 9 & 0.14, 0.14  & 21, 37 \\
2MASS J15450901$+$3555271    &    L7.5    &    N    &    N    & \nodata & [3.6], [4.5] & 14, 7 & 0.8, 0.9  & 26, 37 \\ 
Luhman 16A   &    L7.5    &    N    &    Y    &    $\sim$4.5--5.5 &  $RIzYJHK_{\rm s}$ & several epochs & 0.1--1.5  & 31, 34, 46, 47, 48 \\ 
SDSS J042348.57–041403.5    &     L7.5    &    Y    &   N    & \nodata &  $J$ & 3.57 & 0.02  & 43, 45 \\     
2MASS J01075242$+$0041563    &    L8    &    N    &    Y    &    Irregular &  [3.6], [4.5] & 14, 7 & 0.3, 0.2  & 18, 37 \\
2MASS J16322911$+$1904407    &    L8    &    N    &    Y    &    3.9$\pm$0.2 &  [3.6], [4.5] & 8, 6 & 0.4, 0.3  & 1, 37 \\
2MASSI J0328426$+$230205    &    L9.5    &    N    &    N    &    \nodata & [3.6], [4.5] & 14, 7 & 0.5, 0.6  &  3, 37 \\ 
2MASS J15203974$+$3546210    &    T0    &    N    &    N    &   \nodata &  [3.6], [4.5] & 14, 7 & 0.3, 0.4  &26, 37 \\ 
2MASS J15164306$+$3053443    &    T0.5    &    N    &    Y    &    6.7 & [3.6], [4.5] & 14, 7 & 0.4, 0.5  & 26, 37 \\
Luhman 16B     &    T0.5    &    N    &    Y    &    4.87$\pm$0.01 & $RIzYJHK_{\rm s}$ &several epochs & 0.1--1.5  & 31, 34, 46, 47, 48 \\ 
2MASS J08583467$+$3256275    &    T1    &    N    &    N    &    \nodata &  [3.6], [4.5] & 14, 7 & 0.3, 0.4  & 26, 37 \\ 
2MASS J21392676$+$0220226    &    T2    &    N    &    Y    &    7.72 &  $J$ & 2.5 & 0.05  & 43, 45 \\
SDSS J075840.33$+$324723.4    &    T2    &    N    &    Y    &    4.9$\pm$0.2 & $J$ & 3.54 & 0.02  &  43, 45 \\
2MASS J120956131$-$10040081    &    T2$+$T7.5    &    N    &    N    &    \nodata & [3.6], [4.5] & 14, 7 & 0.7, 0.7  &  37, 43 \\ 
SIMP J013656.5$+$093347.3    &    T2.5    &    N    &    Y    &    2.3895$\pm$0.0005 & $J$ & 4 nights & 0.5  & 25, 43 \\
SIMP J162918.41$+$033537.0    &    T3    &    N    &    Y    &    6.9$\pm$2.4 & $J$ & 4.03 & 0.01  & 43, 45 \\
SDSS J102109.69$-$030420.1    &    T4    &    N    &    N    &    \nodata & $J$ & 3.17 & 0.7  & 23, 33, 43  \\ 
2MASS J05591914–1404488    &    T4.5$/$T5    &    N    &    N    &    \nodata & $J$ & 3.52 & 0.07  &  43, 45 \\ 
2MASSI J2254188$+$312349    &    T5    &    N    &    N    &    \nodata &  [3.6], [4.5] & 14, 7 & 0.7, 0.7  & 37, 43 \\ 
2MASS J12255432$-$2739466    &    T6    &    N    &    N    &    \nodata &  $J$ & 2.85 & 0.7  & 33, 43 \\ 
2MASSI J1534498$-$295227    &    T6    &    N    &    N    &    \nodata &  $J$ & 3.88 & 0.7  & 33, 43 \\ 
SDSSp J162414.37$+$002915.6    &    T6$/$T6    &    N    &    N    & \nodata & $I$ & 4 nights & 4.4  &  35, 43 \\     
2MASSW J1047539$+$212423    &    T7    &    Y    &    N    &    1.77$^{\rm ll}$ &  $J$ & 9.9$^{\rm l}$ & 1.1  &39, 43 \\
2MASS J12171110$-$0311131    &    T7    &    N    &    Y    &    \nodata & $J$ & 3.2 & 0.07  & 43, 45 \\ 
2MASS J12373919$+$6526148    &    T7    &    Y    &    Y    &    \nodata &  $J$ & 2.5 & 2.3  & 6, 43 \\ 
2MASSI J0415195$-$093506    &    T8    &    N    &    Y    &    \nodata &  $J$ & 3.79 & 0.01  & 43, 45 \\ 
2MASS J07271824$+$1710012    &    T7$/$T8    &    N    &    N    & \nodata &  $J$ & 5.0 & 1.1  & 32, 43 \\    
\enddata
\begin{minipage}{\textwidth}
\tablenotetext{}{{\bf Notes}: $^{\rm a}$ Total duration for 11 different observing epochs of $\sim$6 h each with an average photometric error is 0.3$\%$; $^{\rm b}$ Total duration for 3 different observing epochs with an average photometric error of 0.5$\%$; $^{\rm c}$ $R$- and $I$-band data taken from references 15 and 7, respectively; $^{\rm d}$ Four runs with lengths of 2--3.8 h each and a typical photometric error of 1.3 \%; $^{\rm e}$ Five runs with lengths of  1.7--6.8 h each and a typical photometric error of 0.5 \%; $^{\rm f}$ Five runs with lengths of 1.0--5 h each; $^{\rm g}$ $I$-band data consists of 4 runs of 1.6--5.3 h each; $^{\rm h}$ Total time for 3 runs of 4.4--6.2 h each; $^{\rm i}$ Total time for 3 runs of 3.8 h each; $^{\rm j}$ Total time for 3 runs of 4.2--5.3 h each; $^{\rm k}$ Total time for 3 runs of 4.5--5.4 h each; $^{\rm l}$ Total time for 5 runs using different telescopes with a typical photometric error of 1.1$\%$;  $^{\rm ll}$ Rotation period derived from radio observations \citep{2015ApJ...808..189W}.}
\tablenotetext{}{{\bf References.}  (1) \citet{1999ApJ...519..802K}; (2) \citet{2000AJ....119..928F}; (3) \citet{2000AJ....120..447K}; (4) \citet{2000AJ....120.1085G}; (5) \citet{2001AA...367..218B}; (6) \citet{2002AJ....123.2744B}; (7) \citet{2002ApJ...577..433G}; (8) \citet{2002MNRAS.332..361C}; (9) \citet{2003MNRAS.339..477B}; (10) \citet{2003ApJ...594..510B}; (11) \citet{2003MNRAS.346..473K}; (12) \citet{2003ApJ...583..451M}; (13) \citet{2003AJ....125..343L}; (14) \citet{2004MNRAS.354..378K}; (15) \citet{2005ApJ...619L.183M}; (16) \citet{2005MNRAS.360.1132K}; (17) \citet{2006MNRAS.370.1208L}; (18) \citet{2007AJ....133.2258S}; (19) \citet{2007AJ....133.1633M}; (20) \citet{2007ApJ...659..675R}; (21) \citet{2008ApJ...684.1390R}; (22) \citet{2008PASP..120..860B}; (23) \citet{2008ApJ...689.1295K}; (24) \citet{2009AJ....137.3345C}; (25) \citet{2009ApJ...701.1534A}; (26) \citet{2010AJ....139.1808S}; (27) \citet{2011ApJ...739...49B}; (28) \citet{2012ApJ...746...23M}; (29) \citet{2013MNRAS.428.2824K}; (30) \citet{2013ApJ...779..101H}; (31) \citet{2013AA...555L...5G}; (32) \citet{2013ApJ...767...61G}; (33) \citet{2014AA...566A.111W}; (34) \citet{2014ApJ...790...90F}; (35) \citet{2015ApJ...801..104H}; (36) \citet{2015AJ....149..158S}; (37) \citet{2015ApJ...799..154M}; (38) \citet{2015MNRAS.453.1484R}; (39) \citet{2015MNRAS.448.3775R}; (40) \citet{2015ApJ...813..104G}; (41) \citet{2015ApJS..220...18B}; (42) \citet{2015ApJ...808..189W}; (43) \citet{2016ApJ...826...73P}; (44) \citet{2014ApJ...797..120R}; (45) \citet{2014ApJ...793...75R}; (46) \citet{2013ApJ...778L..10B}; (47) \citet{2015ApJ...812..163B}; (48) \citet{2016ApJ...825...90K}.}
\end{minipage}
\end{deluxetable*}

\bibliographystyle{aasjournal}
\bibliography{miles-paez}

\end{document}